# Upholding the diffraction limit in the focusing of light and sound

A. A. Maznev[a] and O. B. Wright[b]

[a] Department of Chemistry, Massachusetts Institute of Technology, Cambridge, MA 02139, USA

[b] Division of Applied Physics, Faculty of Engineering, Hokkaido University, Sapporo 060-8628, Japan

**Abstract**

The concept of the diffraction limit put forth by Ernst Abbe and others has been an important guiding principle limiting our ability to tightly focus classical waves, such as light and sound, in the far field. In the past decade, numerous reports have described focusing or imaging with light and sound 'below the diffraction limit'. We argue that the diffraction limit defined in a reasonable way, for example in terms of the upper bound on the wave numbers corresponding to the spatial Fourier components of the intensity profile, or in terms of the spot size into which at least 50% of the incident power can be focused, still stands unbroken to this day. We review experimental observations of 'subwavelength' or 'sub-diffraction-limit' focusing, which can be principally broken down into three broad categories: (i) 'super-resolution', i.e. the technique based on the modification of the pupil of the optical system to reduce the width of the central maximum in the intensity distribution at the expense of increasing side bands; (ii) solid immersion lenses, making use of metamaterials with a high effective index; (iii) concentration of intensity by a subwavelength structure such as an antenna. Even though a lot of interesting work has been done along these lines, none of the hitherto performed experiments violated the sensibly defined diffraction limit.

## 1. Introduction

The concept of the diffraction limit was formulated towards the end of the 19th century in the works of Abbe, Rayleigh and others on the resolution of optical instruments [1, 2]. Diffraction impedes the resolution of small features in microscopy and objects of small angular separation in astronomy; it also limits our ability to focus waves into a small spot, which is important, for example, in photolithography. Not surprisingly, scientists and engineers have been trying to do better than the diffraction limit seems to prescribe, and, in many instances, succeeded. For example, in fluorescence microscopy it is now possible to resolve features much smaller than half the optical wavelength $\lambda$ [3]. Likewise, in photolithography, it is possible to print features much smaller than $\lambda/2$ [4]. Interestingly, in both cases the propagation of light remains strictly within the constraints imposed by the diffraction limit. What made these developments possible is that, for example, in photolithography, one is ultimately interested in producing small features in the photoresist rather than in the optical intensity pattern. Consequently, various intrinsically nonlinear techniques such as double exposure [4] can be used to fabricate photoresist features much smaller than the smallest possible far-field focused laser spot. Likewise, in microscopy with the manipulation of fluorescence–based detection, such as stimulated emission depletion or on-and-off stochastic switching of fluorophore molecules [3], one can resolve subwavelength features of an object even if such features in the optical intensity distribution cannot be resolved. It is also possible to resolve subwavelength features using near-field optical methods, in which case structures with subwavelength dimensions such as needles, tapered fibers or optical antennas are used to confine or scatter light [5,6].

An important development occurred in the year 2000, when Pendry [7] extended the concepts of focusing using materials with negative permittivity $\varepsilon$ and permeability $\mu$ developed by Veselago [8] to show that focusing of light to a subwavelength spot (in theory, to a point) is possible with a slab of a double-negative material $\varepsilon = \mu = -1$ (i.e., refractive index $n=-1$). Pendry's 'superlens' seemed to break the diffraction limit in earnest. It turned out, however, that in any practical situation, i.e. in the presence of losses, the superlens only works at subwavelength distances in the near-field of the source [9]. Indeed, the effect discovered by Pendry should be classed as a near-field effect, as it relies on evanescent rather than propagating waves; only in the idealized system consisting of a block of lossless negative index material with $\varepsilon = \mu = -1$ considered by Pendry does it persist in the far field. Thus, in any real situation this superlens does not overcome the diffraction limit, which applies to far-field propagating waves. In contrast, focusing of rays via negative refraction proposed by Veselago [8] does work in the far field but is subject to conventional limitations imposed by diffraction. The near-field 'superlens' and the far-field 'Veselago lens' are two different phenomena: the latter only requires that both $\varepsilon$ and $\mu$ be negative and their product be unity, resulting in $n=-1$ (assuming that the second medium is vacuum with $n=1$) [8]. Furthermore, the group velocity opposite to the phase velocity, which leads to 'Veselago focusing' via negative refraction, is encountered in systems without double negativity [10]. Pendry's ideal superlens effect, on the other hand, is specific to the case of a lossless double-negative medium with $\varepsilon = \mu = -1$. The same considerations apply to the focusing of acoustic waves by a double-negative (i.e. negative effective density and elastic modulus) slab of material [11].

It is instructive to observe that the superlens effect can be achieved without a negative-index material if one uses an array of deeply subwavelength sources (for example, acoustic transducers or radio-frequency antennas) to recreate the same field as would be produced by a negative-index slab. However, in order to transmit subwavelength features to an image plane located in the far field using evanescent waves, one would need to drive the transducers at unrealistically high amplitudes. For example, a transducer array can easily generate a field pattern with a period of $\lambda/2$ (yielding an intensity pattern with a period of $\lambda/4$) comprised of two evanescent waves given by $\exp(i\omega t \pm i2kx - \sqrt{3}kz)$, where $k$ is the acoustic or optical wavenumber and $z$ is the distance from the array. The resulting intensity pattern will have a period of $\lambda/4$ but will fade away as $\exp(-2\sqrt{3}kz)$. At a distance of $\lambda$ from the array it will decay by a huge factor of $\sim 3 \times 10^9$. Thus even though subwavelength imaging with evanescent waves is theoretically possible at any distance from the source, it is only practical in the near field at sub-wavelength distances.

Despite the limitations of the superlens restricting its effect to the near field, the excitement generated by Pendry's paper led to extensive work on subwavelength focusing and imaging, and in the ensuing years multiple groups reported 'breaking' the diffraction limit in the far field in both optics and acoustics: sub-diffraction-limited focusing or imaging was observed with metamaterials and without metamaterials, with negative refraction and without negative refraction, with Helmholtz resonators in acoustics and with Maxwell's fish eye lenses in optics [12-17]. The general mood was expressed by a commentary in Nature Materials entitled "What diffraction limit?" [18], implying that the diffraction limit was all but irrelevant. In this paper, we will argue that the diffraction limit has not become irrelevant—in fact it is particularly useful in analyzing reports of sub-diffraction-limited resolution and elucidating the origin of the observed phenomena. In the following, we will concentrate on focusing because it conceptually simplifies the discussion: focusing deals exclusively with wave propagation, whereas the issue of imaging resolution involves the interaction of wave fields with the object being imaged. Focusing is understood here as a far-field phenomenon occurring over distances larger than the wavelength; below we will show that distinguishing between far-field focusing and near-field 'hot spots' is essential for a correct interpretation of experimental results.

## 2. What is the diffraction limit?

To begin with, we need to define what the diffraction limit is. Reports of sub-diffraction-limit focusing or imaging typically do not provide a precise definition. Rather, it is often simply stated that the diffraction limit corresponds to a dimension of $\lambda/2$, typically with a reference to the Rayleigh criterion. Recall that the Rayleigh criterion pertains to resolving two Airy disks produced by two incoherent point sources, and posits that the two disks can be just resolved when the first minimum of one coincides with the maximum of the other, which, in application to a microscope yields a resolution limit of $0.61\lambda/N$,[1] where $N$ is the numerical aperture, i.e. the sine of the half angle of the focused ray cone [19]. However, the Rayleigh criterion has never been considered a hard limit. Born and Wolf, in their classic text Principles of Optics [19], explain that the Rayleigh criterion is "…appropriate to direct visual observations. With other methods of detection (e.g. photometric) the presence of two objects of much smaller angular separation than indicated by Rayleigh's criterion may often be revealed." Indeed, if any number of photons is available for the measurement, there is *no* fundamental limit to how well one can resolve two point sources, since it is possible to make use of curve fitting to arbitrary precision (however, there are obvious practical limitations related to the finite measurement time and other factors such as imperfections in the optical system, atmospheric turbulence, etc.)

The diffraction limit of $\lambda/2$ is also often cited with respect to the full width at half maximum (FWHM) of the intensity profile of the focal spot. This notion originates from the FWHM of the central maximum of the Airy disk, which is indeed close to $0.5\lambda/N$, as shown in Fig. 1(a). However, Rayleigh himself knew very well that one can do better than that [20]: for example, the central maximum of a diffraction pattern produced by an annular aperture, shown in Fig. 1(b), has a FWHM of $\sim 0.36\lambda/N$. A narrower central maximum is achieved at the expense of larger sidebands. Moreover, in 1952 Toraldo di Francia [21] showed that the aperture can be so modified as to make the width of the central maximum arbitrarily small. Again, this is achieved at the expense of increasing sidebands, which become much larger than the central maximum, as shown in Fig. 1(c). Thus, as noted by Born and Wolf [19], the resolution is ultimately limited only by the amount of light available.

---

[1] The resolution depends on the numerical aperture $N$ because the incident beam can be decomposed into a Fourier sum of plane waves travelling in different directions, so that the superposition and interference of these waves gives rise to the image. The in-plane component of the wave vector of each incident plane wave in the image plane will increase with the angle of incidence of the wave. Therefore to achieve the highest resolution one requires the widest range of incident angles possible, i.e. up to 90°, in which case $N=1$.

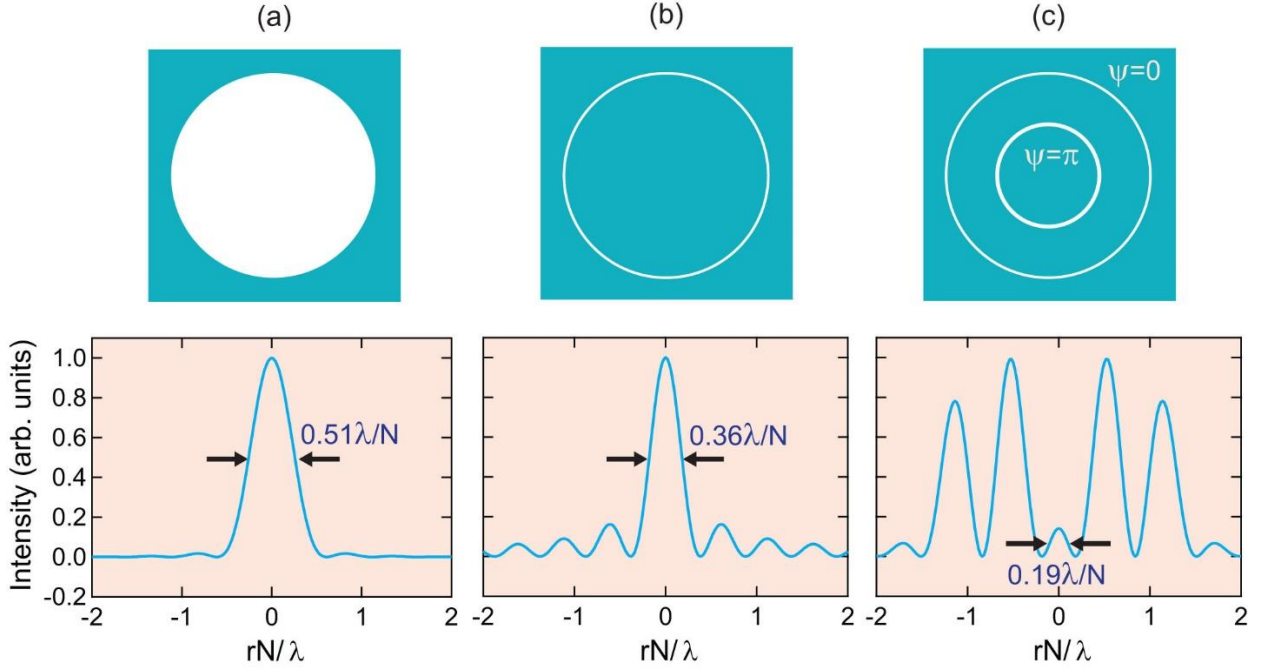

Figure 1. 'Super-resolution' focusing: a narrow central spot is achieved at the expense of increasing sidebands. Intensity profiles in (a) an Airy disk produced by a circular aperture; (b) diffraction pattern produced by an annular aperture; (c) diffraction pattern produced by a simple apodized aperture, in which the inner ring is two times smaller in radius and 1.5 times thicker compared to the outer one, and is covered by a phase-shift mask that flips the phase by $\pi$.

The question thus arises, is it possible to define the diffraction limit in a sensible way? We believe that it is, and would like to offer two definitions concerning the diffraction limit for focusing.

(i) *The smallest period of the in-plane spatial Fourier components of the energy density distribution in the image plane cannot be less than $\lambda/2$, where $\lambda$ is the wavelength in the medium (or 'effective medium' for a metamaterial).*

This follows since the in-plane components of the wave vector of a propagating wave cannot exceed the wave vector magnitude, i.e. $2\pi/\lambda$; consequently, the Fourier-transform of the energy density, which is proportional to the field squared, cannot contain wavenumbers exceeding $4\pi/\lambda$. If we take into account the numerical aperture $N=\sin\theta$, where $\theta$ ($\leq 90°$) is the half-angle of the outlying cone of rays exiting the focusing system, then the smallest period of the in-plane spatial Fourier components of the energy density distribution in the image plane cannot be less than $\lambda/2N$. (Here we do not include the refractive index in the definition of $N$, but use the wavelength in the medium $\lambda=\lambda_0/n$ rather than the wavelength in vacuum $\lambda_0$—because in acoustics there is no unique way to define a refractive index due to the lack of a universal reference speed of sound analogous to the speed of light in vacuum $c$.)

The above definition is similar in essence to Abbe's definition of the imaging resolution of a microscope for a periodic structure [1]. However, in discussing the resolution in microscopy one should also consider illumination and its interaction with the object under study [22-24]. If the illumination contains large wave vector components, then high wave vector spatial Fourier components of the object can be revealed even if the imaging system only lets through low wave vector spatial Fourier components. This can be achieved, for example, by illuminating the object by evanescent waves from a high-index substrate [25], or from a subwavelength structure placed in the proximity of the object [26]. The same principle is used in illumination-mode near-field scanning optical microscopy with near-field illumination and far-field detection [27]. Alternatively, evanescent waves from an object incident on a subwavelength grating will produce propagating waves carrying information about high spatial Fourier components of the object. (This scheme has been used, for example, in the 'far-field superlens' [28].) Nonlinear interaction of the illuminating light field with the object, as for example in fluorescence microscopy [3], further complicates finding a definition of the resolution limit. However, these complexities do not make the diffraction limit irrelevant: it remains true that spatial Fourier components of the fluorescence intensity with a period less than $\lambda/2$ cannot be imaged, even if it is possible to resolve much finer Fourier components in the object itself.

In application to focusing, one may wish to have a more practical definition of the diffraction limit in terms of one's ability to concentrate light or sound to a small spot. As discussed above, the FWHM of the focal spot is not a good measure of the energy concentration. A more relevant measure would be, for example, the smallest spot size containing 50% of the total energy incident on the image plane. In the Airy disk case, for example, 50% of the energy is contained in a spot of $0.535\lambda/N$ in diameter. Maximizing the energy contained within a circle of a given diameter by shaping the incident field (by modifying the transmission function of the exit pupil of the focusing lens) is a constrained extremum problem which has been considered by Lansraux and Boivin [29]. Their numerical results indicate that that for a circle of $\lambda/2N$ in diameter one can do only marginally better compared to an unobstructed aperture yielding an Airy disk. We thus propose an alternative definition of the diffraction limit in focusing:

(ii) *More than 50% of the total energy cannot be focused into a spot smaller than $\sim\lambda/2N$ in diameter.*

Another popular measure of the energy concentration is the Strehl ratio, i.e. the ratio of the maximum intensity at the focal point to the total power [30]. Luneburg [31] has shown that this ratio is the highest for an unobstructed aperture producing an Airy disk, in which case it is equal to $\pi N^2/\lambda^2$ (although this is strictly accurate only for small values of $N$). In practice, the Strehl ratio is typically made dimensionless by taking the value for the unobstructed aperture as unity [19, 30]. Any modification of the aperture used to achieve 'super-resolution', as shown in Fig. 1, only makes the Strehl ratio smaller. The requirement that the dimensionless Strehl ratio should not exceed unity can be used as an alternative to definition (ii).

For either definition to be sensible, focusing should take place in the far field. A subwavelength antenna can yield a deeply subwavelength peak of the optical or acoustic field, but that will have nothing to do with focusing. Hence we define the 'far field' by the requirement that the distance from the location where the intensity is measured to any subwavelength object should be large compared to $\lambda$. (The same applies to the distance to any high-index medium, otherwise one should use the value of $\lambda$ in that medium.) This restriction appears to exclude focusing in metamaterials which by definition are made of subwavelength elements. However, metamaterials can be included as long as we average the fields over a distance much larger than the size of the structural features, just as we average optical fields in conventional materials over a distance much greater than, for example, a crystal lattice constant. The proposed definitions are formulated for monochromatic fields; however, they remain valid for non-monochromatic fields [16, 32, 33] if the shortest wavelength of the spectrum is taken as $\lambda$, with the first definition (i) applying both to the time-integrated intensity, i.e., the energy density distribution in the image plane, and to the instantaneous pattern of the square of the field in this plane at any particular time.

## 3. Has sub-diffraction-limited focusing been demonstrated?

Let us now apply our proposed definition (either version) to the reported instances of sub-diffraction-limited focusing/imaging.[2] We find that such reports generally fall into three categories, schematically illustrated in Fig. 2.

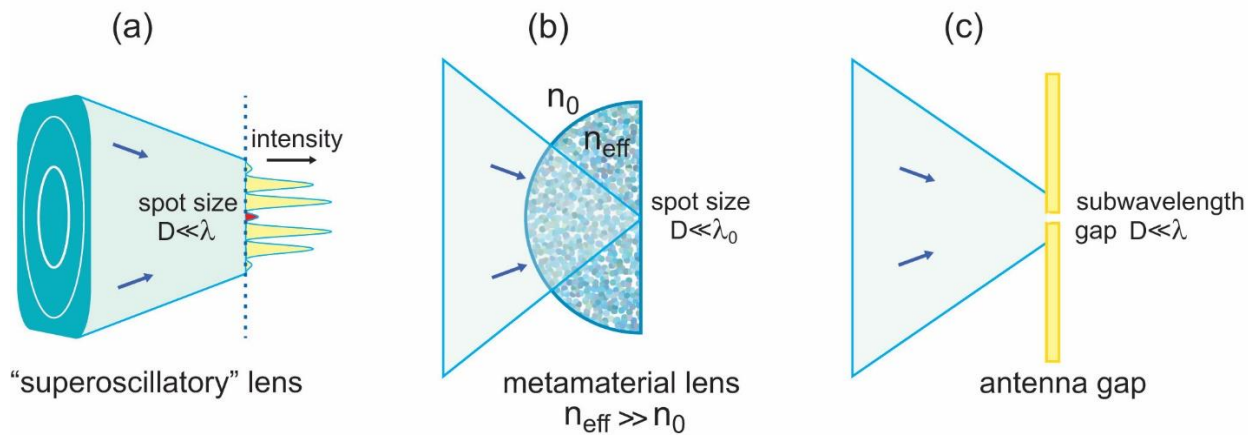

Figure 2. Three categories of reported instances of sub-diffraction-limited focusing. (a) Super-resolution: improving resolution by modifying the pupil of the optical system (apodization). (b) Solid immersion lenses with metamaterials: hyperlenses and metalenses. (c) Near-field hot spots: use of subwavelength features such as antennas.

### 3.1. Super-resolution/superoscillations

Improving resolution by modifying the pupil of the optical system (apodization) following the path broken by Toraldo di Francia has been traditionally referred to as 'super-resolution' [34, 35]. Recently, efforts in this direction have been intensified [14, 36, 37, 38], taking advantage of the progress achieved in nanofabrication. A reduction of the FWHM of the central spot below $\lambda/2$ at the expense of large sidebands has also been observed in Veselago-type acoustic focusing by phononic-crystal structures without apodization [15, 33]. In either case, no contradiction to the proposed definitions of the diffraction limit arises. In particular, even deeply subwavelength hot spots do not contradict the $\lambda/2$ limit on the period of the highest spatial Fourier component of the intensity distribution.[3] Functions that can locally oscillate faster than their highest Fourier-component have been investigated in quantum mechanics, and the phenomenon is termed 'superoscillations' [39]. In 2006, Berry and Popescu [40] proposed to use this effect for super-resolution focusing, apparently without knowing of the prior work by Toraldo di Francia and others. Since then, the term 'superoscillations' is commonly used to describe this phenomenon in the propagation of classical waves such as light and sound [14, 33, 36, 37].

---

[2] We will concentrate on experimental reports as theory/simulations papers are too numerous to be comprehensively reviewed here.

*3.2. Solid immersion lenses with metamaterials*

A large body of work on far-field sub-diffraction-limited focusing and imaging with 'hyperlenses' and 'metalenses' made of optical and acoustic metamaterials has been reviewed by Lu and Liu [41]. The principal question that should be asked here is what the optical/acoustic wavelength $\lambda$ is in the metamaterial medium. In fact, one finds that statements of 'subwavelength' resolution are invariably based on a comparison with the wavelength $\lambda_0$ either in vacuum (in optics) or in the surrounding conventional medium (in acoustics). If the wavelength in the metamaterial is considered, no violations of the diffraction limit are found. A case in point is the 'hyperlens' [12, 13, 42] made of an electromagnetic hyperbolic metamaterial. In an ideal lossless hyperbolic medium, the dispersion relation is given by $(k_x^2+k_y^2)/\varepsilon_\parallel + k_z^2/\varepsilon_\perp = \omega^2/c^2$, where dielectric tensor components $\varepsilon_\parallel$ and $\varepsilon_\perp$ have opposite signs. The isofrequency surfaces are hyperbolic, hence at any given frequency the wave vector magnitude is unbounded close to the asymptotic directions, which implies an arbitrarily small wavelength (although in practice the wave vector magnitude is limited by losses [43]).

Another example is the focusing of sound above an array of Helmholtz resonators (actually soda cans) [16], which form a locally-resonant metamaterial [44], to a spot as small as $\lambda/25$, where $\lambda$ is the acoustic wavelength in air. In the lossless effective-medium limit, the wave vector of the guided mode in the metamaterial medium diverges as the frequency approaches the Helmholtz resonance from below, hence the wavelength becomes much smaller than the wavelength in air [45]. This is not entirely evident in experiments using broad-band time-reversal [16], but becomes clear using monochromatic waves [45]. Time-reversal allows an impressive degree of control over focusing [32], but it does not enable sub-diffraction-limited resolution. The sharp focusing observed in time-reversal experiments is not subwavelength with respect to the wavelength in the metamaterial [45, 46], and can be achieved without time reversal [45].

A peculiar kind of metamaterial can be produced by stacking non-interacting waveguides together [47]. A rigid pipe filled with fluid or gas supports acoustic waveguide modes with wavelengths much greater than the diameter of the pipe. If we stack many such pipes and make every pipe carry one pixel of an image [47], we get what appears to be deeply subwavelength imaging. And if we use expanding pipes to make a 'magnifying hyperlens' [48], then deeply-subwavelength focusing also appears possible. However, as soon as the stack of pipes is formally treated as an effective medium, 'subwavelength' resolution disappears. Indeed, the isofrequency surfaces of such effective medium are flat, as shown in Fig. 3 (or almost flat in the case of weakly interacting pipes [48]). Consequently, the in-plane component of the wave vector can be arbitrarily large, just as in the above-mentioned hyberbolic material case. Obviously it is those large in-plane wave vector components that encode the image information.

---

[3] The formation of deeply subwavelength hot spots can be illustrated by the following simple example: consider a maximum of the field in a field pattern in the image plane. Now add a spatially uniform out-of-phase field (meaning spatially uniform within the image plane which is easily produced by a plane wave) to make only the very top of this maximum stick out above the zero level. This will produce an intensity maximum as narrow as we wish, albeit only containing a tiny fraction of the total energy.

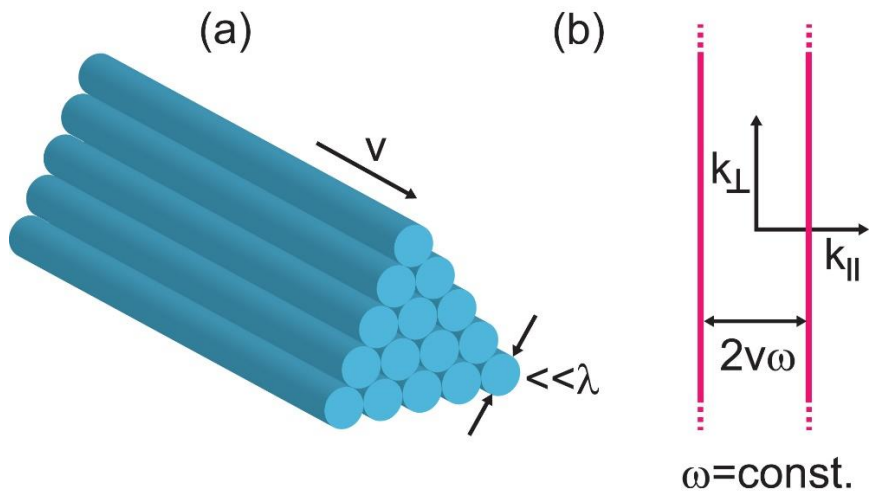

Figure 3. (a) A stack of identical rigid pipes each of diameter much smaller than the acoustic wavelength $\lambda$ and filled with liquid or gas can be formally considered as a metamaterial yielding flat isofrequency surfaces such as the one shown by red lines in (b). The sound velocity in the pipes is denoted by $v$.

Thus 'subwavelength' focusing with metamaterials is similar in essence to focusing with a solid immersion lens made of a natural material with a high refractive index [49]. This is not to deny that many studies aimed at achieving sub-diffraction-limited resolution with metamaterials are interesting in their own right and that a number of unique designs demonstrated with metamaterials would not be possible with natural materials, the hyperlens being a prime example of such a design.[4]

### 3.3. Near-field 'hot spots'

It is well known that a structure with subwavelength features such as an antenna can produce a deeply subwavelength 'hot spot' of an optical or acoustic field. For example, Fig. 4 shows an acoustic intensity profile above a Helmholtz resonator in the form of a soda can at a frequency of 410 Hz, close to the resonance frequency of the can [45]. A narrow peak with a FWHM of about $\lambda/35$ is observed just above the opening of the can. This is a near-field effect, but if the can is placed at the focal point of a focusing system, an appearance of a subwavelength focal spot is created. Thus in any instance of reported sub-diffraction-limited focusing one needs to check for the presence of subwavelength structures in the proximity of the 'focal spot'.

One study that resulted in an extensive debate was that of perfect imaging with a Maxwell's fish-eye lens [17, 50-60]. The caveat was that 'perfect imaging' required a point drain placed at the focal point. An interested reader is referred to recent studies [59, 60] involving some of the co-authors of the original experimental report [17]; these two papers show quite convincingly that a Maxwell's fish-eye lens does not yield sub-diffraction-limited imaging.

---

[4] Even though natural hyperbolic materials do exist [43], they cannot be easily shaped into a hyperlens.

Another example is focusing with a time-reversal mirror combined with a time-reversed point source [61]. The latter yields a subwavelength peak that disappears once the point source is removed. In the acoustic Helmholtz-resonator array experiment [16], the 'focal spot' appears particularly small owing to a combination of the small wavelength in the metamaterial medium near the local resonance, as discussed in the previous sub-section, and the near-field effect illustrated in Fig. 4, i.e. the concentration of the acoustic intensity at the opening of a soda can [45]. We should remember that metamaterials are made of subwavelength elements. The concept of a metamaterial implies, strictly speaking, use of the effective medium approach in which fields are averaged over a distance large compared to the size of the 'unit cell', whereas the fine structure of the field will inevitably contain sharp subwavelength features that have nothing to do with far-field focusing.

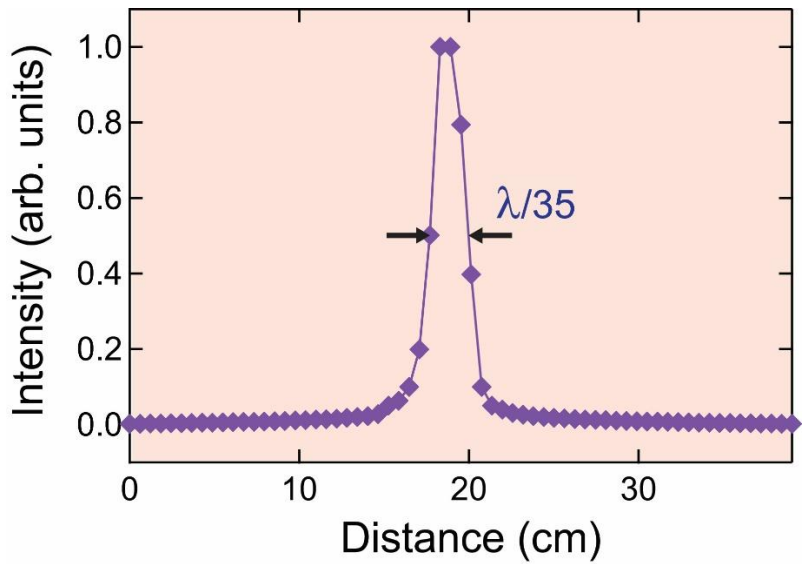

Figure 4. Acoustic intensity profile above a single soda can at 410 Hz (based on experimental data from Ref. 45) yields a narrow peak above the opening of the can.

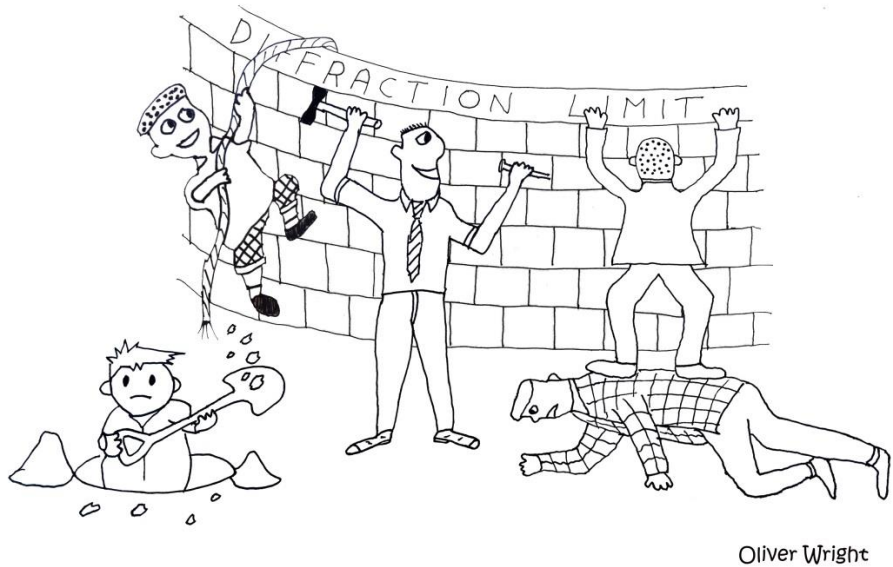

Figure 5. The diffraction limit stands firm.

## 4. Conclusion

The desire to overcome the diffraction limit has motivated a lot of great work, as evidenced by the 2014 Nobel Prize in chemistry, and we hope that more great work in this area is still to come. Yet, the concept of the diffraction limit stands firm: far from becoming irrelevant, it is in fact even more useful in analyzing recent experiments involving complex materials, negative refraction, time-reversal, etc. What is the wavelength in the metamaterial medium? Do we see near-field effects from subwavelength structures involved? Is subwavelength focusing achieved at the expense of large sidebands? Answering these questions will help guide analysis of experimental results. Precisely defining what we mean by saying 'diffraction limit' is more than just a question of semantics. Defining things clearly helps us understand what exactly is achieved when new results are reported and better appreciate the limitations and opportunities for using light or sound to probe small length scales. We hope that our attempt to provide a working definition of the diffraction limit will stimulate a productive discussion in the research community that will contribute to greater understanding of the issues concerned with focusing in optics, acoustics, and other fields involving wave propagation.

## Acknowledgments

The authors thank Jean-Jacques Greffet, Keith Nelson, and Vincent Laude for helpful comments on the manuscript. The contribution by A.A.M. was supported by the NSF Grant No. CHE-1111557, USA. The contribution by O.B.W. was supported by Grants-in-Aid for Scientific Research from the Ministry of Education, Culture, Sports, Science and Technology (MEXT), Japan.

[57] U. Leonhardt, T. G. Philbin, Reply to 'comment on 'perfect imaging with positive refraction in three dimensions", Phys. Rev. A 82 (2010) 057802.

[58] R. Merlin, Maxwell's fish-eye lens and the mirage of perfect imaging, J. Opt. 13 (2011) 024017.

[59] T. Tyc, A. Danner, Resolution of Maxwell's fisheye with an optimal active drain, New J. Phys. 16 (2014) 063001.

[60] S. He, F. Sun, S. Guo, S. Zhong, L. Lan, W. Jiang, Y. Ma, T. Wu, Can Maxwell's Fish Eye lens really give perfect imaging? Part III. A careful reconsideration of the 'Evidence for subwavelength imaging with positive refraction', Progress in Electromagnetics Research 152, (2015) 1-15.

[61] J. de Rosny, M. Fink, Overcoming the diffraction limit in wave physics using a time-reversal mirror and a novel acoustic sink, Phys. Rev. Lett. 89 (2002) 124301.
14